# Toughness and Strength of Nanocrystalline Graphene

**Ashivni Shekhawat**[1,2,3] **and Robert O. Ritchie**[1,2]

[1]Materials Sciences Division, Lawrence Berkeley National Laboratory, Berkeley, CA 94720

[2]Department of Materials Science & Engineering, University of California, Berkeley, CA 94720

[3]Miller Institute for Basic Research in Science, Berkeley, CA 94720

**Abstract:** Pristine monocrystalline graphene is claimed to be the strongest material known with remarkable mechanical and electrical properties. However, graphene made with scalable fabrication techniques is polycrystalline and contains inherent nano-scale line and point defects – grain boundaries and grain-boundary triple junctions – that lead to significant statistical fluctuations in toughness and strength. These fluctuations become particularly pronounced for nanocrystalline graphene where the density of defects is high. Here we use large-scale simulation and continuum modeling to show that the statistical variation in toughness and strength can be understood with 'weakest-link' statistics. We develop the first statistical theory of toughness in polycrystalline graphene, and elucidate the nano-scale origins of the grain-size dependence of its strength and toughness. Our results should lead to more reliable graphene device design, and provide a framework to interpret experimental results in a broad class of 2D materials.

**Subject terms:** Materials science; Stochastic fracture mechanics; Nanoscale simulations; 2D materials

The high strength of graphene combined with its exceptional electronic [1, 2], optical [3], and thermal properties [4] has made it an ideal material for many fascinating applications, including flexible electronic displays [5], corrosion-resistant coatings [6], biological devices [7, 8], and many more [9]. While each of these applications exploits a different key property of graphene, they all implicitly depend on its exceptional mechanical properties for structural reliability [10]. However, such mechanical reliability of graphene is impacted by atomic defects in its structure. While the effect of relatively simple defects, such as isolated dislocation cores or a few special grain boundaries, on the strength of graphene is understood[11-17], the statistical fluctuations in strength and toughness due to the randomness in polycrystalline nanostructure remains largely unexplored.

Strength and toughness are arguably the two most important properties of a structural material. Whereas strength is generally a function of the material's resistance to deformation, toughness represents its resistance to fracture. In most materials, these properties tend to be mutually exclusive [18]. There are conflicting experimental reports whether the strength of polycrystalline graphene is actually a function of grain size[19-21] making the role of simulation and theory more critical. Understanding these statistical fluctuations has become important in light of the fact that graphene synthesized with chemical vapor deposition (CVD) is polycrystalline, and this method is being used to manufacture more than 300,000 m$^2$ of graphene annually [22, 23]. In this work we develop an understanding of the grain-size dependent statistical fluctuations in strength and toughness of polycrystalline graphene by using a combination of weakest-link statistics, continuum elastic theory, and large-scale molecular dynamics (MD) simulations.

It is well established that the strength and toughness of polycrystalline solids is strongly influenced by their granular structure. For instance, nanocrystalline metals are invariably significantly harder and much less ductile than their microcrystalline counterparts. This is due to the well-known "Hall-Petch effect", wherein the motion of dislocations, and thereby plastic flow, is impeded by the presence of grain boundaries [24]. On the other hand, dislocations are typically not mobile in brittle materials, such as ceramics (and graphene), and thus the Hall-Petch effect is not observed in these materials. An entirely different mechanism leads to grain-size dependent strength in brittle ceramics where the length of the typical extrinsic crack-like flaw (inclusion/porosity) relative to the grain size determines the characteristic strength [25, 26]. In contrast to typical bulk brittle materials such as ceramics, graphene can be fabricated in a much cleaner and controlled environment, thus making the presence of large extrinsic defects unlikely [22]. In the absence of such extrinsic defects, the fluctuations in strength must arise from intrinsic atomic defects inherent to the granular nanostructure. The traditional theories developed for brittle ceramics with large extrinsic flaws are thus not applicable for strength fluctuations due to these intrinsic defects. In graphene these defects are grain boundaries (GBs) and triple junctions (TJs). Although a GB is an interface between crystalline regions of different lattice orientations and a TJ is the intersection of three such interfaces, in graphene GBs and TJs are typically composed of pentagon-heptagon defects, also known as 5-7 defects (Figure 1)[27-30]. These defects involve significant residual stresses and act like stress concentrators. Here we find that while the strength of a brittle polycrystalline graphene sheet is dictated by the weakest flaw in the entire sheet, its toughness is conversely influenced by the GB nearest to the crack tip. We develop theories to capture these two markedly different mechanisms. We believe that the theoretical framework developed here will be applicable to a large class of emerging 2D materials.

## Results

### Statistical Distribution of Strength

We use molecular dynamics (MD) simulations to gain insights into the statistical distribution of polycrystalline strengths[31-33]. We simulate fracture in over 19,000 polycrystalline graphene sheets with random grain shapes and orientations at several different combinations of sheet size, grain size and strain rate. The details of our simulations can be found in the Methods section and Supplementary Material. Figure 2A shows a schematic representation of one such simulation. Figures 2B and 2C show the snapshots of a polycrystalline simulation at zero stress, and at peak stress just before global failure, respectively. Residual stress at GB and TJ defects can be seen in Figure 2B; the length-scale associated with these stresses is sub-nanometer, showing that the fluctuations are truly an atomic-scale phenomena. Figure 2C shows that fracture originates at a grain boundary and then progresses through rest of the polycrystal. This observation is generic and fracture always originates at a GB or TJ in the several thousand polycrystals that we have simulated. This is not surprising since the interior of the grains are defect free and have no residual stresses. Once fracture originates, the incipient crack can extend in an intragranular or intergranular manner depending on the details of the grain orientation and loading direction.

Traditionally the statistical distribution of fracture strengths in brittle materials is understood in terms of Weibull theory[34-37]. According to Weibull, the survival probability of brittle materials, defined as the probability that a volume $V$ of the material survives at a stress σ, is of the form $S(\sigma) = \exp\left(-V\left(\frac{\sigma-\sigma_0'}{v'}\right)^m\right)$ where the lower-bound strength, $\sigma_0'$, and $v'$ are, respectively, the location and scale parameters, and $m$ is the Weibull modulus. These parameters are generally evaluated by data fitting; however, in our case, fitting this from simulation data for each combination of grain size, strain rate and system size would give a different value for these parameters, and thus would not provide any physical insight into the dependence of these parameters on the nanostructural features such as grain size. As our goal is to gain such insight, we derive an expression for the survival probability that explicitly accounts for the dependence on the physical atomistic and structural variables of interest.

The specific details of our derivation can be found in the Supplementary Material; we present here the main result. We consider a polycrystalline graphene sheet of linear size $L$ with a linear grain size $\mu$. The sheet is loaded uniaxially at a constant strain rate $\dot{\epsilon}$ at a fixed temperature of 300 K. Thus, the stress at time $t$ is given by $\sigma(t) = \left(\frac{Y}{1-\eta^2}\right)\dot{\epsilon}t$, where $Y$ is the Young's modulus and $\eta$ is the Poisson's ratio. Our main result for polycrystalline strength gives the following expression for the survival probability of the sheet loaded at strain rate $\dot{\epsilon}$ up to a stress $\sigma$:

$$S(\sigma|L,\mu,\dot{\epsilon}) = e^{-\frac{L^2\dot{\epsilon}_0}{\mu^2\dot{\epsilon}}\left(\frac{\sigma-\sigma_0}{v}\right)^m}, \qquad [1]$$

where $\dot{\epsilon}_0$ is a reference strain rate for normalization, and $\sigma_0, v$ are the rescaled location and scale parameters. As opposed to the usual Weibull form, in Eq. [1] the effect of structural parameters such as system size, grain size and loading rate is captured by a single non-dimensional parameter $L^2\dot{\epsilon}_0/\mu^2\dot{\epsilon}$, while the rescaled parameters $\sigma_0, v, m$ are not affected by these details and are true material properties. We arrive at this particularly simple result by assuming that the individual defects in the graphene sheet are non-interacting, and that the loading is slow enough so that a thermal quasi-equilibrium is achieved. The dependence on $L^2/\mu^2$ is due to the fact that fracture initiates at GB and TJ defects, and the number of such defects in a graphene sheet scales as $\sim L^2/\mu^2$. The factor $\dot{\epsilon}_0/\dot{\epsilon}$ is due to the fact that a lower strain rate gives more opportunity for thermal nucleation of fracture at the defects. Our model is a 'weakest-link' model, since we assume that the graphene sheet fractures as soon as its weakest GB/TJ defect becomes unstable. One particularly interesting outcome of our analysis is that the strength of graphene is a function of the ratio $L^2/\mu^2$, thus there is no one well-defined value of strength at a given grain size. As we shall show, this behavior is in contrast to the toughness, which is well defined for given a grain size. Finally, it is worth noting that we use a powerful theorem from extreme value statistics as a crucial step in our derivation [38, 39].

We test our theoretical result with large-scale MD simulations. We perform extensive statistical sampling for 24 different combinations of the parameters $L, \mu, \dot{\epsilon}$. Figure 3A shows the variation of the survival probability with grain size, while Figure 3B shows the effect of strain rate. A joint fit of the survival probability for the entire dataset using the form given in Eq. [1] is obtained from the maximum likelihood estimator. This fit yields a Weibull modulus of $m = 10.7$, a location parameter of $\sigma_0 = 19.5$ GPa, and a scale parameter of $\nu = 53.2$ GPa. Figure 3C shows that with these values all the data for the various combinations of $L, \mu, \dot{\epsilon}$ collapses on single line, validating our theoretical result. Eq. [1] further predicts that the mean fracture strength of graphene should scale as:

$$\langle \sigma \rangle = \sigma_0 + \nu \left(\frac{\dot{\epsilon}}{\dot{\epsilon}_0}\right)^{\frac{1}{m}} \left(\frac{\mu}{L}\right)^{\frac{2}{m}} \Gamma\left(1 + \frac{1}{m}\right), \qquad [2]$$

where $\Gamma(\cdot)$ is the Gamma function. Figure 3D presents a validation of this relation. Note that after $\sigma_0, \nu, m$ have been obtained by fitting the survival probability to Eq. [1], there are no more free parameters, and thus the graph in Figure 3D contains no free parameters. The scaling of the mean strength with grain size in Eq. [2] is similar to the result reported by Sha *et al.* [40]; however they did not model the system size and strain-rate dependence.

## Statistical Distribution of Toughness

The strength of a material quantifies its failure in response to a state of large homogeneous stress, although in practice the state of stress is rarely homogeneous; rather, some regions of the material experience much higher stress than others. This is often due to the presence of stress singularities associated with sharp cracks and corners. The fracture toughness quantifies such failure in response to the stress singularity due to the presence of a crack. Not surprisingly, the toughness is one of the most important mechanical properties of a material. For nominally brittle materials the fracture toughness can be evaluated in terms of the critical stress-intensity factor, $K_{I_{cr}}$, which is a measure of the strength of the stress singularity at the crack tip (the stress intensity specifically quantifies the magnitude of the elastic stress and displacement fields at the crack tip). Griffith's criterion establishes a lower bound for the critical stress-intensity factor as $K_{I_{cr}} \geq \sqrt{2\gamma Y}$, where $\gamma$ is the energy required to create the fresh crack surface. However, the Griffith theory does not account for statistical fluctuations in $K_{I_{cr}}$ due to variations in the local nanostructure of GBs and TJs. Here we develop a theory to account for these factors.

We use MD simulations to explore the statistical fluctuations in fracture toughness and its dependence on grain size in graphene. Figure 4A shows a schematic representation of our simulations. We simulate the initial advance of semi-infinite cracks in graphene polycrystals by imposing suitable fixed boundary conditions away from the crack tip, and evolving the atoms near the crack tip with canonical NVT[**] dynamics; specific details can be found in the Methods section. As shown in Figure 4B, the stress singularity at the crack tip interacts with the GB and TJ defects in the immediate vicinity of the tip. The critical stress-intensity factor needed to initiate crack advance depends on the local nanostructure; as an arbitrary crack would experience a random sampling of this nanostructure, there is significant statistical spread in the values of $K_{I_{cr}}$ and thus it is meaningful to define a grain-size dependent survival probability, $S(K_I|\mu)$, as the probability that a polycrystal with grain size $\mu$ survives an applied stress-intensity factor $K_I$. Figure 5B shows the numerically obtained survival probability from our MD analysis on polycrystals with different grain sizes. The figure clearly shows that although there is significant spread in the observations, there is a strong dependence on grain size.

---

[**] NVT is Number Volume Temperature, which is refers to a simulation at fixed number of atoms, fixed volume and fixed temperature.

In general, we have observed that cracking progresses by breaking the bonds associated with the pentagon-heptagon defects near the crack tip. In fact, due to lack of plasticity, the stress concentration at the crack tip in graphene is so strong that the first bond to break is almost always within 10 Å of the crack tip. Because the polycrystalline morphology is generated randomly, as the grain size is increased it becomes more likely that there are no defects near the crack tip. If there are no GBs or TJs near the tip, then the toughness is simply given by toughness of monocrystal that contains the crack tip. Since the monocrystal will have an arbitrary orientation, a part of the statistical spread in our simulation results is due to the variation in fracture toughness of pristine graphene with orientation. For the AIREBO empirical potential [41, 42] used in our MD simulations, we find that $Y = 858$ GPa, and $\gamma$ varies between $5.9 - 6.3$ J/m² for single crystals depending on the orientation, resulting in an orientation-dependent Griffith estimate of fracture toughness between $3.2 - 3.3$ MPa·m$^{1/2}$ for monocrystalline graphene. In practice the observed values of the toughness of brittle solids are slightly higher than the Griffith bound due to such effects as lattice trapping and crack roughness. We simulate crack advance in graphene monocrystals with random orientations and find that the resulting distribution of $S_{cr}(K_I)$ is well described by a Gaussian distribution with mean 3.9 MPa·m$^{1/2}$ and standard deviation 0.4 MPa·m$^{1/2}$. This somewhat accounts for the behavior in the upper tail (large $K_I$) of Figure 5B, but the interesting grain-size dependent behavior for smaller values of $K_I$ needs to be explored further. We note that the trend of increasing toughness with grain size is opposite to the trend reported in reference [43]; presumably due to the fact that we are measuring in the value of $K_{Ic}$ strictly a crack-initiation toughness (at instability) while reference [43] reports a propagation toughness, and their simulations seem to have non-physical crack bridging with single atom carbon chains (a non-physical phenomenon typical of the AIREBO potential, which we avoid in this work by the design of our simulations).

We now develop a theory to explain the grain-size dependence of $S(K_I|\mu)$ in graphene. In order to maintain theoretical tractability we do not account for loading-rate dependence, and all our simulations of initial crack advance are performed in the quasi-static limit. The stress field in the region of $K$-dominance is known from linear-elastic fracture mechanics (LEFM). For ductile materials with brittle inclusions the statistical variation in toughness can be obtained by integrating the Weibull-type expression of stress survival probability (our Eq. [1]) over the LEFM crack-tip stress field [44, 45]. However, this technique implicitly assumes that a large population of defects is sampled by each crack tip; while this assumption is valid for ductile materials with large crack-tip process zones, it is clearly violated for brittle materials like graphene where the crack tip samples only a few defects within a few angstroms of the tip. Thus, a new approach is needed. It can be argued that the effect of TJs on the distribution of $K_I$ should be minimal, since the probability of finding a TJ near a crack tip is much smaller than the probability of finding a GB nearby. Figure 5A shows a schematic representation of our model. We consider an arbitrary GB a distance $l$ ahead of the crack tip, inclined at an angle $\phi$ to the crack. Each point on this GB experiences a different normal stress $\sigma_n$ that can be calculated from LEFM to be $\sigma_n = \frac{K_I}{\sqrt{2\pi r}} \cos\left(\frac{\theta}{2}\right)\left(1 + \sin\left(\frac{\theta}{2}\right)\sin\left(\frac{3\theta}{2} - 2\phi\right)\right)$. We define $r^*, \theta^*$ as the point at which the GB experiences the maximum normal stress, $\sigma_n^*$. Note that $\sigma_n^*$ is implicitly a function of $l$ since $r^* = \frac{l \sin \phi}{\sin(\phi - \theta^*)}$. We make the assumption that this GB survives at stress intensity $K_I$ if it can survive the normal stress $\sigma_n^*$. We measure the GB survival probability, $S_{GB}(\sigma_n)$, for applied normal stress by simulating over 4,000 GBs (see Methods and Supplementary Material); the resulting survival probability and probability density are shown in Figure 5C. The polycrystalline survival probability can then be written as:

$$S(K_I|\mu) = \int_{l_c}^{\infty} \int_0^{\pi} S_{GB}\left(\sigma_n^*(\phi, l, K_I)\right) p(\phi) d\phi \, p(l|\mu) dl * S_{cr}(K_I), \qquad [3]$$

where $p(\phi)$ is the probability density of the random GB orientation angle $\phi$, $p(l|\mu)$ is the grain size dependent probability density of the GB distance from crack tip, and $l_c$ is a lower cutoff due to the discreteness of the lattice. This equation essentially means that the polycrystal survives a stress intensity $K_I$ if the nearest GB ahead of the crack tip and the crystal containing the crack survive. Since the polycrystal has random boundary orientations, the distribution of $\phi$ is uniform, giving $p(\phi) = 1/\pi$. The distribution of the GB distance from the crack tip is measured from the randomly generated polycrystals to be a half-Gaussian distribution, $p(l|\mu) = \frac{2e^{-\left(\frac{l^2}{2\alpha^2 \mu^2}\right)}}{\sqrt{2\pi\alpha^2\mu^2}}$, where the parameter $\alpha$ is equal to 0.64. A collapse of the probability density of the measured distance to the nearest GB ahead of the crack tip according to this form is shown in Figure 5D. Finally, we take $l_c$ to be equal to the distance between the centers of the next nearest hexagons in graphene ($= 3a$, where $a$ is the carbon-carbon bond length in graphene). Note that there are no free parameters in Eq. [3]. However, since it is unreasonable to assume that only the nearest GB directly ahead of the crack tip contributes to fracture, we leave $\alpha$ as a free parameter when fitting Eq. [3] to numerical data. The solid lines in Figure 5B show a comparison of the predictions of the grain-size dependent polycrystalline fracture toughness obtained from Eq. [3] with the numerical data obtained from extensive simulation. It should be noted that $\alpha = 0.7$ is obtained from the fitting process, and is very close to the measured value of 0.64. Thus, we are able to derive a formula for the statistical fluctuations in polycrystalline toughness that has only one free parameter, the value of which can also be measured to good accuracy from the polycrystal geometry.

## Discussion

Our prediction of mean toughness of 3-4 MPa·m$^{1/2}$ compares favorably with the only reported experimental measurement of the toughness of graphene [10]. Note that this toughness is not high; specifically, it is three to four times tougher than silicon and pyrolytic carbon [46] yet ~20 to 50% less tough than polycrystalline diamond [47]. We have found that the distribution of toughness, as well as strength, in polycrystalline graphene is strongly dependent on the grain size. However, at larger grain sizes the distribution of toughness becomes less sensitive to grain size. We predict that for grain sizes larger than 256 Å, the toughness will be essentially independent of the grain size. Thus the toughness dependence of grain size is a phenomena limited to nanocrystalline graphene. In contrast, the strength will continue to be grain-size and sample-size dependent, and the experimental results will have to be scaled with the Weibull from (Eq. [1]) to get true material properties. We predict a Weibull modulus of around 10 for polycrystalline graphene. Finally, we note that even though our simulations are for nano-crystalline grains, we expect our results to be valid for much larger micrometer-sized grains, since no new physics is expected to emerge at the intermediate length scales.

What do these results mean in practical terms? The first measurement of the strength of pristine monocrystalline graphene reported an intrinsic strength of about 130 GPa, and a Young's modulus of about 1 TPa. In practical terms these results mean that a soccer ball can be placed on a single sheet of graphene without breaking it. What object can be supported by a corresponding sheet of polycrystalline graphene? It turns out that a soccer ball is much too heavy, and polycrystalline graphene can only support a ping-pong ball! Still remarkable for a one-atom thick material, but not quite as breathtaking anymore.

## Acknowledgements

This work was supported by the Mechanical Behavior of Materials Program at the Lawrence Berkeley National Laboratory, funded by the U.S. Department of Energy, Office of Science, Office of Basic Energy Sciences, Materials Sciences and Engineering Division, under Contract No. DE-AC02-05CH11231. A.S. acknowledges financial support from the Miller Institute for Basic Research in Science, at the University of California, Berkeley, in the form of a Miller Research Fellowship.

## Methods

All numerical simulations were performed with the LAMMPS [48] code by using the AIREBO [41, 42] empirical interatomic potential, with the interaction cutoff parameter set to 1.92 Å [11]. The simulations were conducted in the NVT ensemble at temperature $T$=300 K. The random grain morphology was generated by randomly choosing the required number of grain "centers" and generating the boundaries of the granular domains with a Voronoi construction. The crystalline orientations within each grain were also taken to be random. The atomic positions, particularly at the grain boundaries, were generated by a recently proposed algorithm that yields well annealed GBs [49]. Even though this algorithm was proposed for generation of GBs, it can be used successfully to generate well-annealed polycrystalline samples (see Supplementary Material). All structures used in our strength and toughness simulations were prepared running NVT dynamics at $T$=300 K for 1 picosecond, followed by energy minimization using the conjugate gradient algorithm (allowing out of plane deformations, and allowing the simulation cell dimensions to change to attain zero stress).

For the simulations of polycrystalline strength, a constant strain rate $\dot{\epsilon}$ was imposed using the SLLOD equations [50] as implemented in LAMMPS, and the stress response was measured; the largest observed stress was taken to be the ultimate strength of the sample. We performed simulations for the following 24 combinations of the parameters $L, \mu, \dot{\epsilon}$: (64, 32, 1), (128, 64, 1), (128, 32√2, 1), (128, 64/√3, 1), (128, 32, 1), (128, 32√2, 0.5), (128, 64/√3, 0.5), (128, 32√2, 0.25), (128, 64/√3, 0.25), (256, 128, 1), (256, 64√2, 1), (256, 64, 1), (256, 32√2, 1), (256, 128√2/5, 1), (256, 32, 1), (256, 64√2, 0.5), (256, 32√2, 0.5), (256, 128√2/5, 0.5), (256, 32, 0.5), (256, 64√2, 0.25), (256, 32√2, 0.25), (256, 128√2/5, 0.25), (256, 32, 0.25), (512, 32, 1), where the units of $L, \mu$ are Å, and those of $\dot{\epsilon}$ are $10^9$/s. The number of simulations performed for statistical sampling was $10^4, 10^3, 10^2, 10^2$ for $L =$ 64, 128, 256, 512 Å, respectively.

For polycrystalline toughness we simulated systems of length $L = 256$ Å with a crack tip placed at the center of the simulation box. The atoms outside a radius of 100 Å from the crack tip were fixed according the LEFM $K$-field displacement solution, $u_x = \frac{K_I}{2G}\sqrt{\frac{r}{2\pi}}(\kappa - \cos\theta)\cos\frac{\theta}{2}$, $u_y = \frac{K_I}{2G}\sqrt{\frac{r}{2\pi}}(\kappa - \cos\theta)\sin\frac{\theta}{2}$, $u_z = 0$, where $G$ is the shear modulus, and $\kappa = (3 - \eta)/(1 + \eta)$. The atoms within a radius of 100 Å of the crack tip were evolved with NVT dynamics at temperature $T$=300 K. The radius of 100 Å is chosen because the applied strain beyond this radius is less than 0.02, which is small enough for linear elasticity to be applicable. Also, we note that all bond breaking events occur within a radius of 20 Å, thus the simulation box size is large enough to avoid any finite size effects. The applied stress-intensity factor $K_I$ was increased in increments of 0.1 MPa·m$^{1/2}$. The system was held at each value of $K_I$ for $t = 1$ ps. The critical stress-intensity factor was taken to be the lowest value of $K_I$ for which the crack grows. We simulated initial crack advance in polycrystals with grain size $\mu = 16, 32, 64$ Å. For each grain sizse, crack advance was simulated in 500 polycrystals with random grain morphology.

## References

1. Castro Neto, A. H., Guinea, F., Peres, N. M., Novoselov, K. S. & Geim, A. K., The electronic properties of graphene. *Rev. Mod. Phys.* **81** (1), 109-162 (2009).

2. Novoselov, K. S. *et al.*, Two-dimensional gas of massless Dirac fermions in graphene. *Nature* **438** (7065), 197-200 (2005).

3. Bonaccorso, F., Sun, Z., Hasan, T. & Ferrari, A. C., Graphene photonics and optoelectronics. *Nature Photon.* **4** (9), 611-622 (2010).

4. Balandin, A. A. *et al.*, Superior thermal conductivity of single-layer graphene. *Nano Lett.* **8** (3), 902-907 (2008).

5. Bae, S. *et al.*, Roll-to-roll production of 30-inch graphene films for transparent electrodes. *Nature Nanotech.* **5** (8), 574-578 (2010).

6. Prasai, D., Tuberquia, J. C., Harl, R. R., Jennings, G. K. & Bolotin, K. I., Graphene: Corrosion-inhibiting coating. *ACS Nano* **6** (2), 1102-1108 (2012).

7. Liu, Y., Dong, X. & Chen, P., Biological and chemical sensors based on graphene materials. *Chem. Soc. Rev.* **41** (6), 2283-2307 (2012).

8. Sun, X. *et al.*, Nano-graphene oxide for cellular imaging and drug delivery. *Nano Res.* **1** (3), 203-212 (2008).

9. Geim, A. K. & Novoselov, K. S., The rise of graphene. *Nature Mater.* **6** (3), 183-191 (2007).

10. Zhang, P. *et al.*, Fracture toughness of graphene. *Nature Comm.* **5** (2014).

11. Grantab, R., Shenoy, V. B. & Ruoff, R. S., Anomalous strength characteristics of tilt grain boundaries in graphene. *Science* **330** (6006), 946-948 (2010).

12. Wei, Y. *et al.*, The nature of strength enhancement and weakening by pentagon–heptagon defects in graphene. *Nature Mater.* **11** (9), 759-763 (2012).

13. Khare, R. *et al.*, Coupled quantum mechanical/molecular mechanical modeling of the fracture of defective carbon nanotubes and graphene sheets. *Phys. Rev. B* **75** (7), 075412 (2007).

14. Zhang, T., Li, X., Kadkhodaei, S. & Gao, H., Flaw insensitive fracture in nanocrystalline graphene. *Nano Lett.* **12** (9), 4605-4610 (2012).

15. Wang, M. C., Yan, C., Ma, L., Hu, N. & Chen, M. W., Effect of defects on fracture strength of graphene sheets. *Comp. Mat. Sci.* **54**, 236-239 (2012).

16. Rasool, H. I., Ophus, C., Klug, W. S., Zettl, A. & Gimzewski, J. K., Measurement of the intrinsic strength of crystalline and polycrystalline graphene. *Nature Comm.* **4** (2013).

17. Song, Z., Artyukhov, V. I., Yakobson, B. I. & Xu, Z., Pseudo Hall-Petch strength reduction in polycrystalline graphene. *Nano Lett.* **13** (4), 1829-1833 (2013).

18. Ritchie, R. O., The conflicts between strength and toughness. *Nature Mater.* **10** (11), 817-822 (2011).

19. Lee, G.-H. *et al.*, High-strength chemical-vapor-deposited graphene and grain boundaries. *Science* **340** (6136), 1073-1076 (2013).

20. Ruiz-Vargas, C. S. *et al.*, Softened elastic response and unzipping in chemical vapor deposition graphene membranes. *Nano Lett.* **11** (6), 2259-2263 (2011).

21. Suk, J. W., Mancevski, V., Hao, Y., Liechti, K. M. & Ruoff, R. S., Fracture of polycrystalline graphene membranes by in situ nanoindentation in a scanning electron microscope. *Physica Ptatus Polidi (RRL) – Rapid Research Letters* (2015).

22. Li, X. *et al.*, Large-area synthesis of high-quality and uniform graphene films on copper foils. *Science* **324** (5932), 1312-1314 (2009).

23. Ren, W. & Cheng, H.-M., The global growth of graphene. *Nature Nanotech.* **9** (10), 726-730 (2014).

24. Hall, E. O., The deformation and ageing of mild steel: III Discussion of results. *Proc. Phys. Soc. Sec. B* **64** (9), 747 (1951).

25. Singh, J. P., Virkar, A. V., Shetty, D. K. & Gordon, R. S., Strength-grain size relations in polycrystalline ceramics. *J. Am. Ceram. Soc.* **62** (3), 179 (1979).

26. Rice, R. W., Freiman, S. W. & Mecholsky, J. J., The dependence of strength-controlling fracture energy on the flaw-size to grain-size ratio. *J. Am. Ceram. Soc.* **63** (3-4), 129-136 (1980).

27. Huang, P. Y. *et al.*, Grains and grain boundaries in single-layer graphene atomic patchwork quilts. *Nature* **469** (7330), 389-392 (2011).

28. Yazyev, O. V. & Louie, S. G., Topological defects in graphene: Dislocations and grain boundaries. *Phys. Rev. B* **81** (19), 195420 (2010).

29. Banhart, F., Kotakoski, J. & Krasheninnikov, A. V., Structural defects in graphene. *ACS Nano* **5** (1), 26-41 (2011).

30. Kim, K. *et al.*, Grain boundary mapping in polycrystalline graphene. *ACS Nano* **5** (3), 2142-2146 (2011).

31. Kotakoski, J. & Meyer, J. C., Mechanical properties of polycrystalline graphene based on a realistic atomistic model. *Phys. Rev. B* **85** (19), 195447 (2012).

32. Liu, T.-H., Pao, C.-W. & Chang, C.-C., Effects of dislocation densities and distributions on graphene grain boundary failure strengths from atomistic simulations. *Carbon* **50** (10), 3465-3472 (2012).

33. Sha, Z. D. *et al.*, On the failure load and mechanism of polycrystalline graphene by nanoindentation. *Sci. Rep.* **4** (2014).

34. Epstein, B., Statistical aspects of fracture problems. *J. Appl. Phys.* **19** (2), 140-147 (1948).

35. Weibull, W., A statistical distribution function of wide applicability. *J.Appl. Mech.* **13**, 293-297 (1951).

36. Gulino, R. & Phoenix, S. L., Weibull strength statistics for graphite fibres measured from the break progression in a model graphite/glass/epoxy microcomposite. *J. Mater. Sci.* **26** (11), 3107--3118 (1991).

37. Hui, C. Y., Phoenix, S. L. & Shia, D., The single-filament-composite test: a new statistical theory for estimating the interfacial shear strength and Weibull parameters for fiber strength. *Composites Sci Technol.* **57**, 1707-1725 (1998).

38. de Haan, L., *Extreme Value Theory and Applications* (Springer US, 1994).

39. Bouchaud, J.-P. & Mezard, M., Universality classes for extreme-value statistics. *J. Phys. A: Math. Gen.* **30** (23), 7997 (1997).

40. Sha, Z. *et al.*, Inverse pseudo Hall-Petch relation in polycrystalline graphene. *Scientific Reps* **4** (2014).

41. Brenner, D. W. *et al.*, A second-generation reactive empirical bond order (REBO) potential energy expression for hydrocarbons. *J. Phys.: Cond. Mat.* **14** (4), 783 (2002).

42. Stuart, S. J., Tutein, A. B. & Harrison, J. A., A reactive potential for hydrocarbons with intermolecular interactions. *J. Chem. Phys.* **112** (14), 6472-6486 (2000).

43. Jung, G., Qin, Z. & Buehler, M. J., Molecular mechanics of polycrystalline graphene with enhanced fracture toughness. *Extr Mech Lett* **2**, 52-59 (2015).

44. Lin, T., Evans, A. G. & Ritchie, R. O., A statistical model of brittle fracture by transgranular cleavage. *J. Mech. Phys. Sol.* **34** (5), 477-497 (1986).

45. Lin, T., Evans, A. G. & Ritchie, R. O., Statistical analysis of cleavage fracture ahead of sharp cracks and rounded notches. *Acta Metall.* **34** (11), 2205-2216 (1986).

46. Ritchie, R. O., Fatigue and fracture of pyrolytic carbon: A damage-tolerant approach to structural integrity and life prediction in "Ceramic" heart valve prostheses. *J. Heart Valve. Disease.* **5** (Suppl 1), 9 (1996).

47. Drory, M. D., Dauskardt, R. H., Kant, A. & Ritchie, R. O., Fracture of synthetic diamond. *J. Appl. Phys.* **78** (5), 3083-3088 (1995).

48. Plimpton, S., Fast parallel algorithms for short-range molecular dynamics. *J. Comp. Phys.* **117** (1), 1-19 (1995).

49. Ophus, C., Shekhawat, A., Rasool, H. I. & Zettl, A., Large-scale experimental and theoretical study of graphene grain boundary structures. *Preprint at http://arXiv:1508.00497* (2015).

50. Daivis, P. J. & Todd, B., A simple, direct derivation and proof of the validity of the SLLOD equations of motion for generalized homogeneous flows. *J. Chem. Phys.* **124** (19), 194103 (2006).

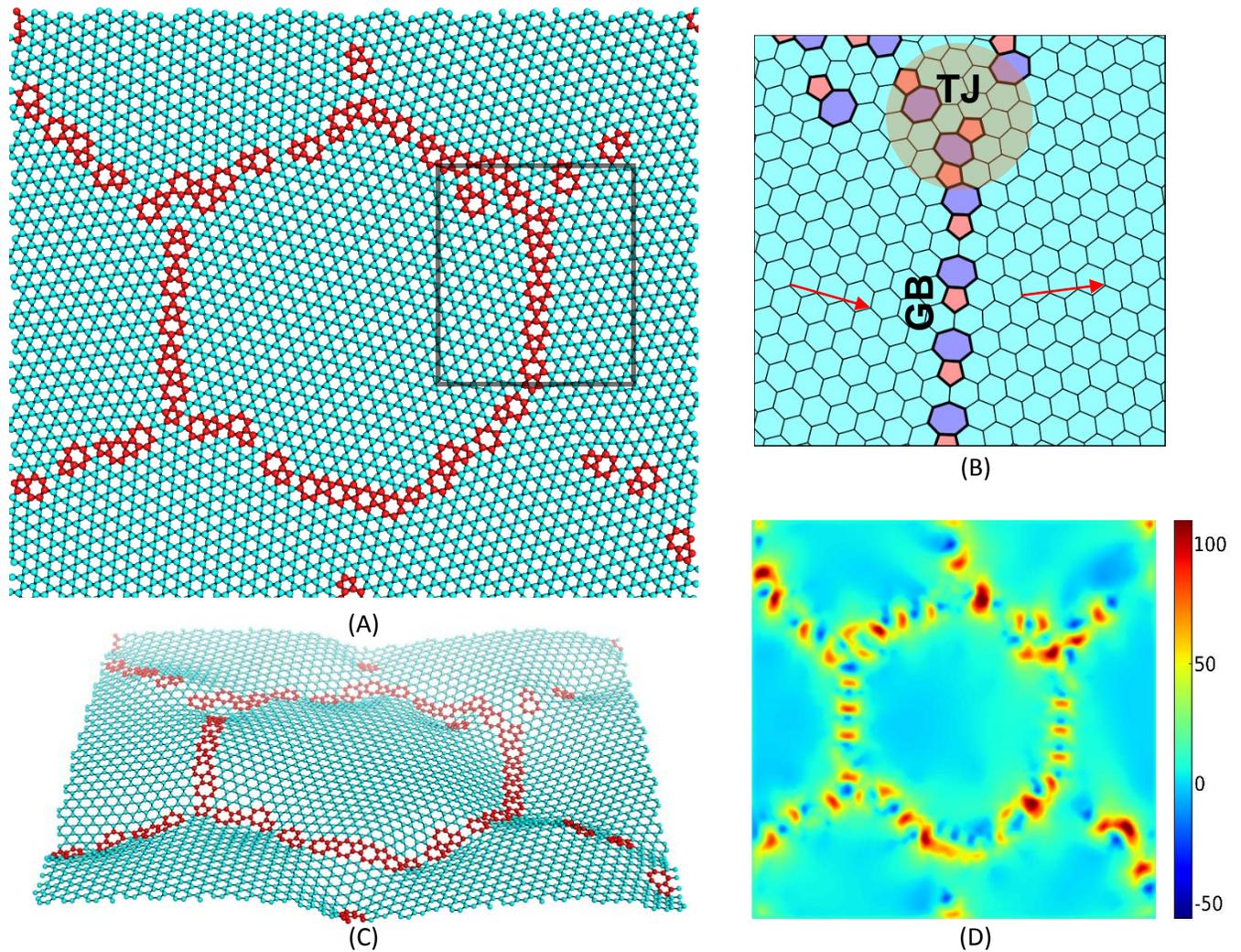

*Figure 1* **(A)** A one atom thick polycrystalline graphene sheet composed of carbon atoms arranged in regular hexagonal rings, except at the grain boundaries (GBs) and triple junctions (TJs). The defected atoms at the GBs and TJs that are part of non-hexagon rings are drawn in red for clear identification. **(B)** A zoom-in of the area marked in (A). The red arrows indicate the orientation of the grains on either side of the GB; the GB itself is composed of rings of 5 (colored pink) and 7 (colored blue) carbon atoms. These are the dislocation cores with the shortest Burgers vectors in graphene, and thus represent low energy configurations of GBs. A TJ formed at the intersection of three GBs is highlighted. **(C)** A perspective view of the graphene sheet showing its 3D structure. **(D)** The principal residual stresses in the graphene sheet are in units of GPa. The grain interiors are defect free, while the GBs and TJs have significant residual stresses, and thus are the weak points where fracture nucleates.

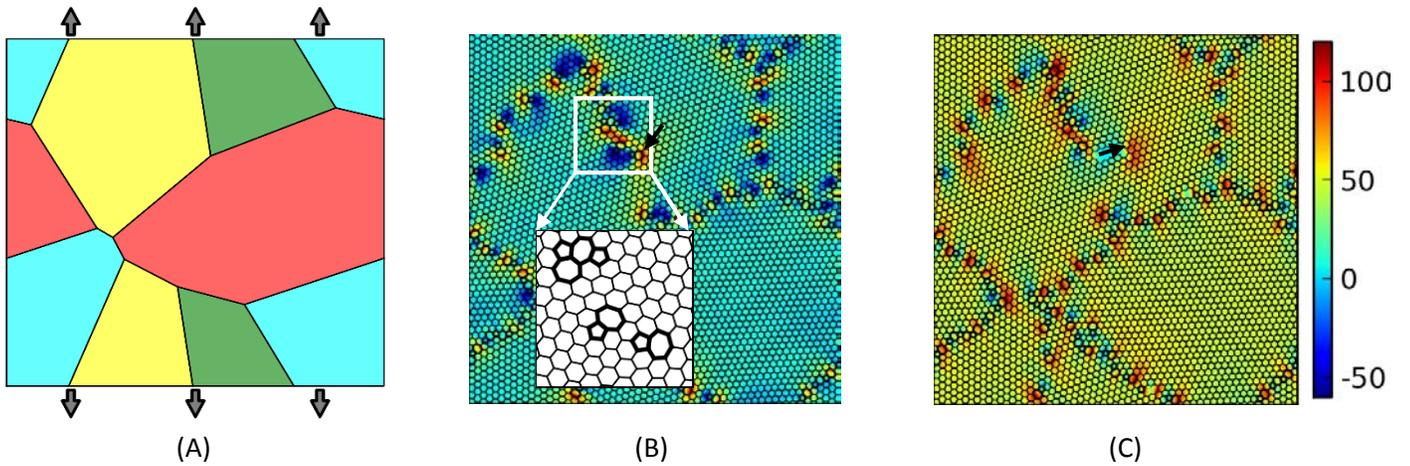

*Figure 2* **(A)** A schematic representation of the fracture simulation of a 2D periodic graphene nanoscale polycrystal. Each color represents a grain within which the lattice orientation is fixed. The polycrystal is loaded at a constant strain rate. **(B)** Residual stress in a polycrystal; the principal stress is plotted in units of GPa. The interior of the grains are stress free while there is significant residual stress at the GBs and TJs. The black arrow points to the large stress concentration at the tail of the pentagon-heptagon defect where fracture ultimately nucleates. The inset shows a zoom-in of the atomic configuration. **(C)** Principal stress just before global fracture. The crack nucleates at the GB defect and extends through the adjacent crystal. Notice that there are a few other incipient cracks that do not go unstable.

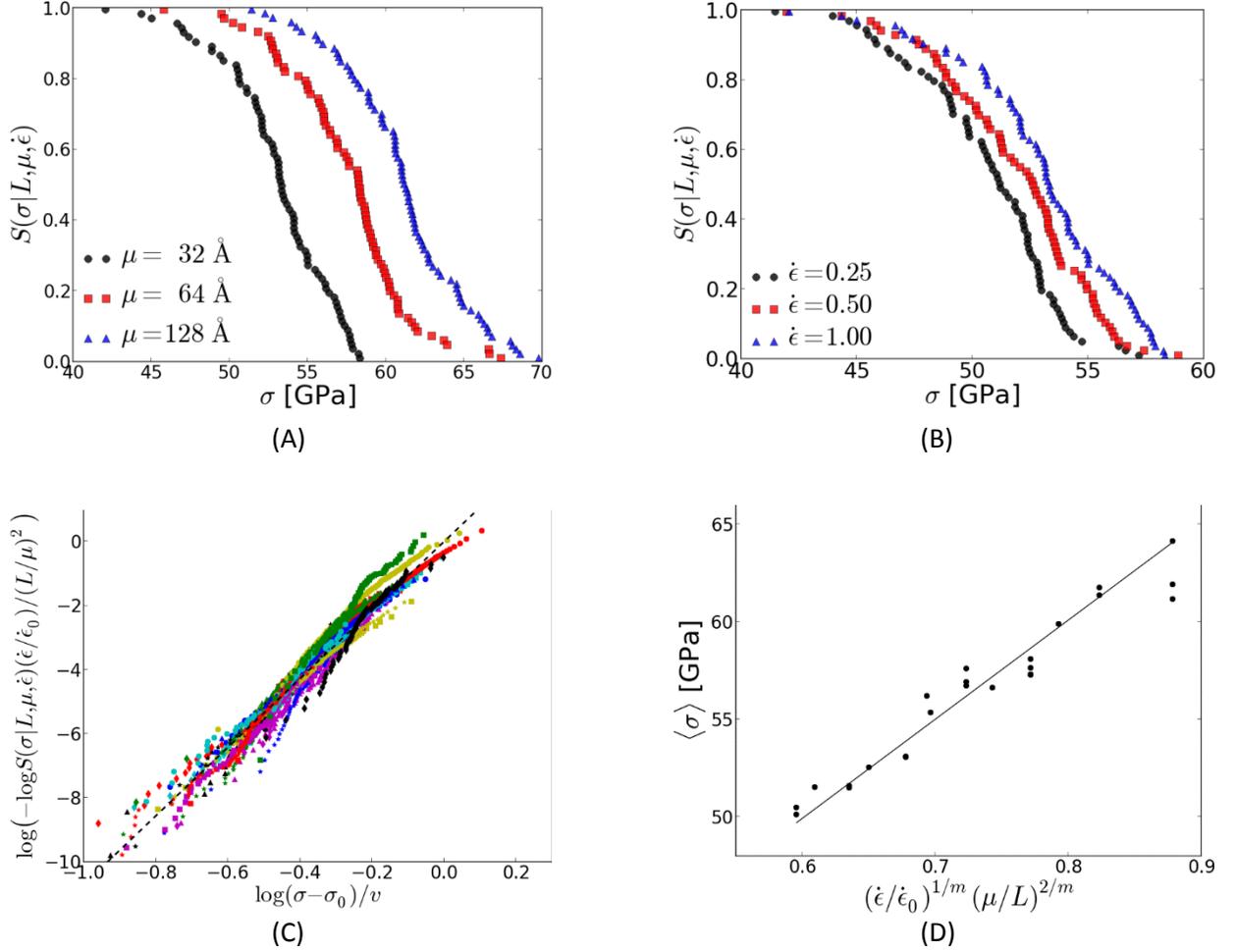

*Figure 3* (**A**) The variation of the survival probability in nanocrystalline graphene with grain size $\mu$ at $L = 256$ Å, $\dot{\epsilon} = 10^9 s^{-1}$. At fixed stress, the smaller grain size has lower survival probability due to a higher density of GB and TJ defects. (**B**) The variation of survival probability with strain rate $\dot{\epsilon}$ (in units of $10^9 s^{-1}$) at $L = 256$ Å, $\mu = 32$ Å. (**C**) A data collapse of the survival probability (obtained from MD simulations) according to Eq. [1] for 24 different combinations of $L, \mu, \dot{\epsilon}$ (see Methods section). The dashed black line is a guide to the eye and shows the prediction of Eq. [1]. The simulation data closely follows the predicted form. (**D**) A collapse of the mean failure stress as predicted by Eq. [2]. Note that this collapse has no free parameters.

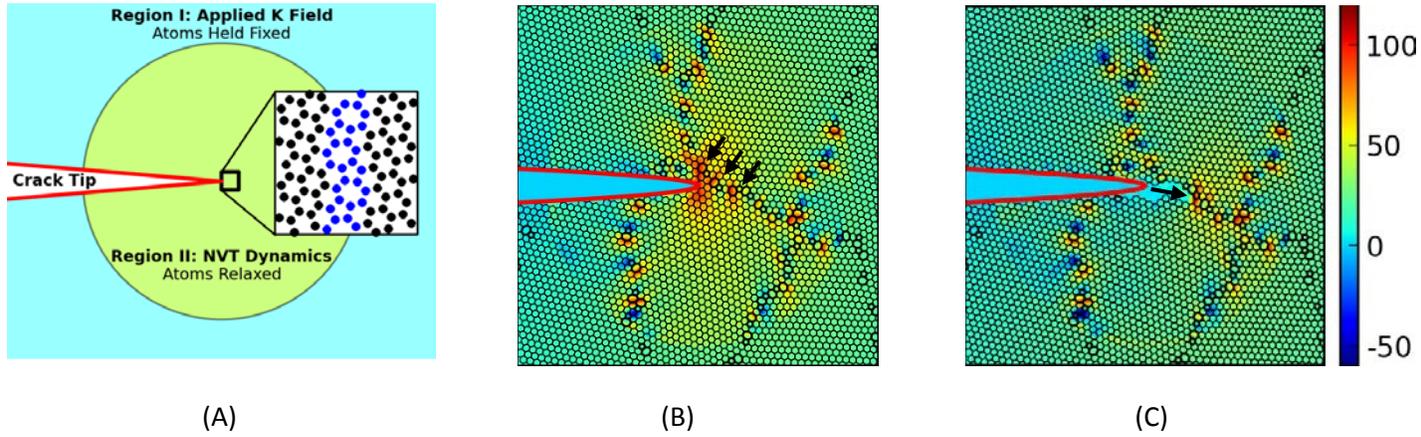

*Figure 4* **(A)** A schematic representation of the simulation of initial crack advance in nanocrystalline graphene. A prescribed $K_I$-field is applied by holding the atoms outside the circular region of radius 100 Å fixed at the positions given by the LEFM solution. The atoms inside the circular region are allowed to relax with NVT dynamics. Initial crack advance occurs when the stress intensity $K_I$ is sufficiently large. **(B)** The stress field in a polycrystal where the atoms outside a 100 Å radius from the crack tip are held fixed at the LEFM displacement field for $K_I = 4$ MPa·m$^{1/2}$. The black arrows show the enhancement of the crack-tip stress concentration by the GB defects. **(C)** The crack extends along the GB as indicated by the black arrow. The stress concentration is relieved after crack extension.

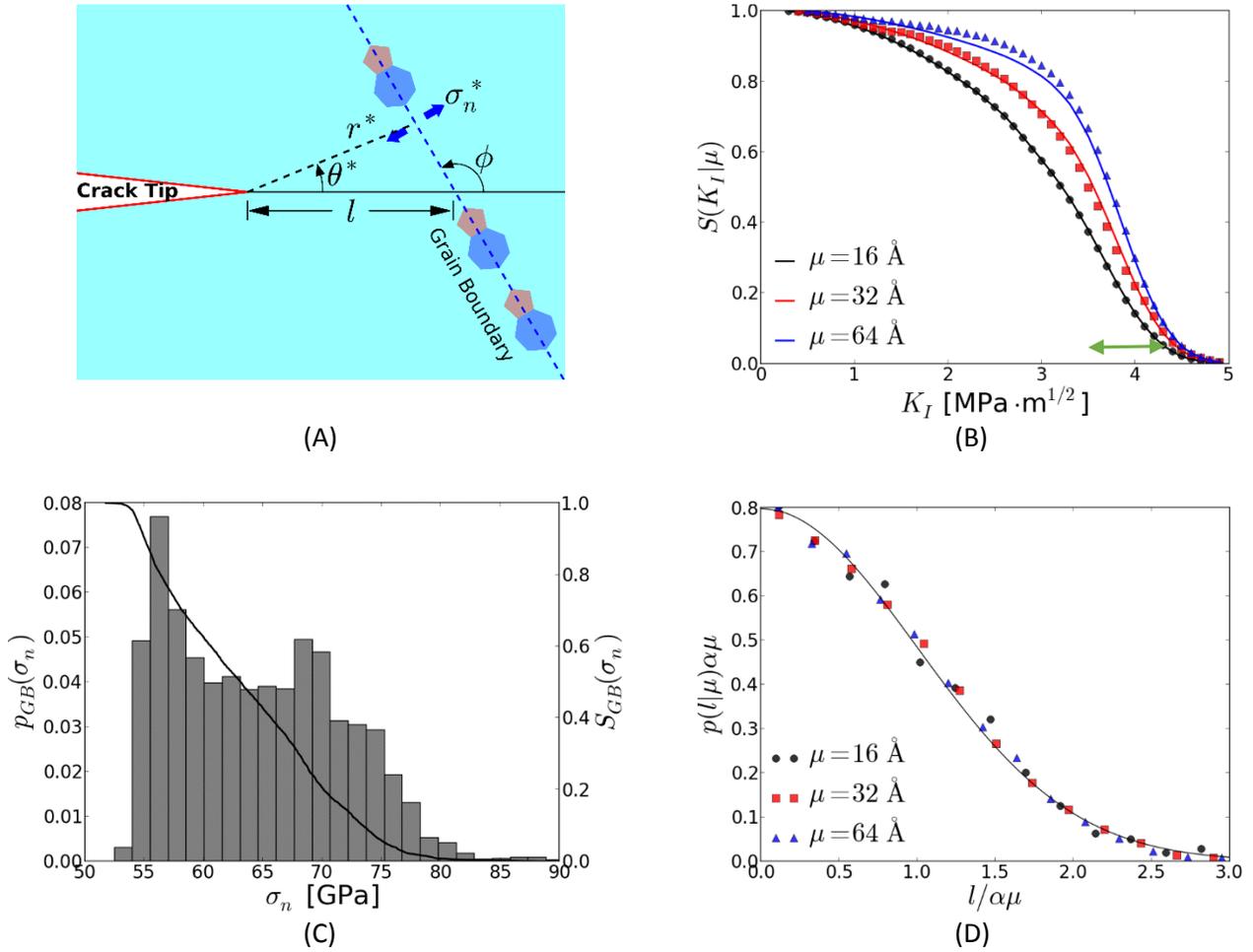

*Figure 5* **(A)** A schematic representation of our model for the toughness of nanocrystalline graphene. A GB a distance $l$ ahead of the crack tip experiences the maximum normal stress at the point $(r^*, \theta^*)$. We assume that the polycrystal survives if the nearest GB and the crystal containing the crack survives the stress due to the crack. **(B)** A comparison of the numerically measured (from MD) survival probability for various grain sizes $\mu$ with the predictions of Eq. [3]. The theoretical predictions are shown in the solid lines, while the MD data are shown in the symbols. Agreement between theory and simulation is evident. The toughness range for monocrystalline samples is indicated by the green line (the spread is due to change in crack orientation with respect to crystal axis). It is evident that crack-trapping due to GBs is minimal, and in most cases GBs lower toughness by facilitating crack advance. **(C)** The grain-boundary survival probability and failure probability density under applied normal stress measured from simulation of over 4,000 GBs. **(D)** The distribution of the distance of the nearest GB ahead of the crack tip measured from the randomly generated polycrystals with different grain sizes. The solid line is a plot of the half-Gaussian fit.